\documentstyle[epsf]{aipproc}
\def\dir{}   

\def\fighadroprod
{
\begin{figure}[t]
 \begin{center}
 \leavevmode
 \epsfxsize=0.45\hsize \epsfbox{\dir 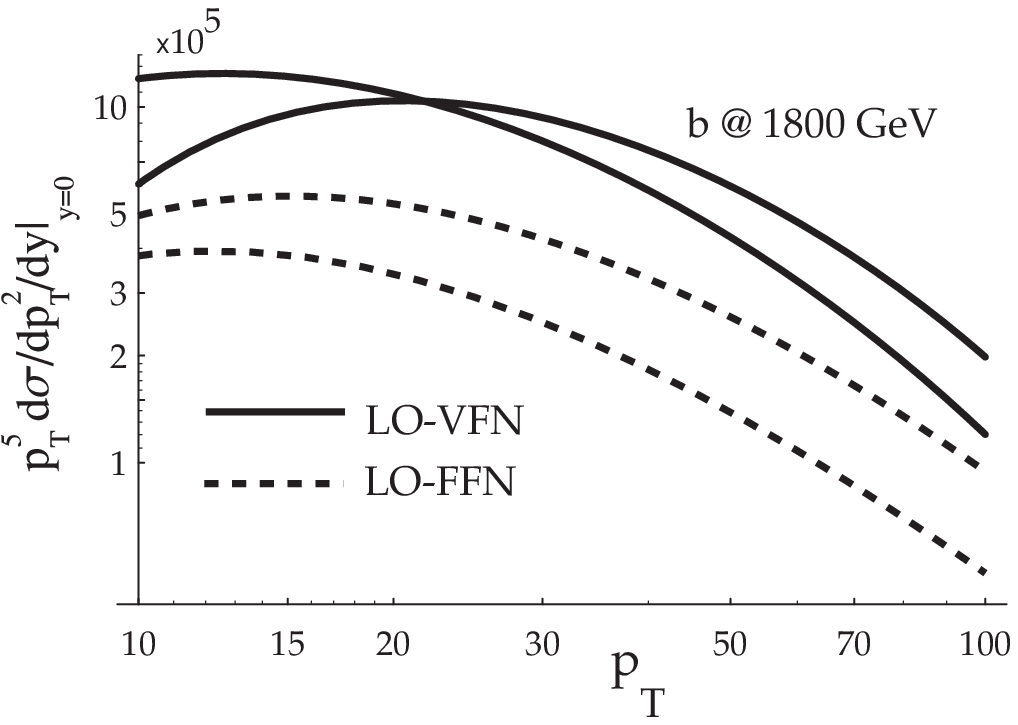} \hfill
 \epsfxsize=0.45\hsize \epsfbox{\dir 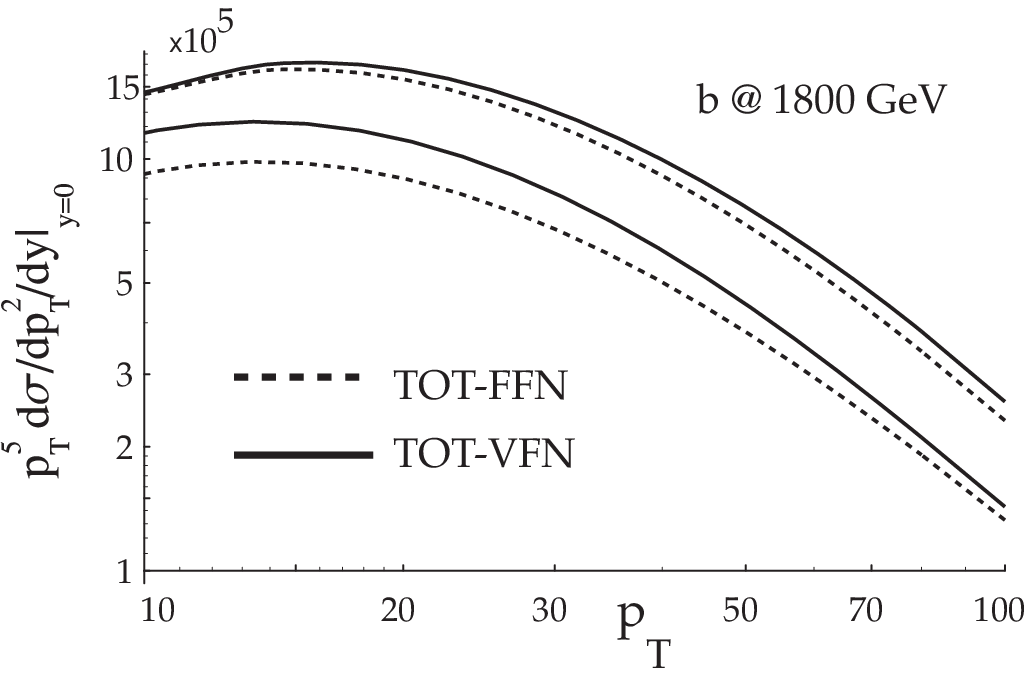}
 \end{center}
 \tightenlines
 \caption{\noindent 
The scaled cross section  ($nb$-GeV$^3$) vs.\ $p_T$ for the 
a) LO-FFN  and LO-VFN  contributions, and 
b) TOT-FFN and TOT-VFN contributions.
For each contribution, we choose $\mu= [M_T/2, 2 M_T]$, 
with $M_T=\sqrt{m_H^2+p_T^2}$,
to gauge the $\mu$-variation. From Ref.~\protect\cite{ost}.
 }
 \label{fig:hadroprod}
\end{figure}
}
\def\figxslimita
{
\begin{figure}[t]
 \epsfxsize=0.60\hsize  \centerline{\epsfbox{\dir 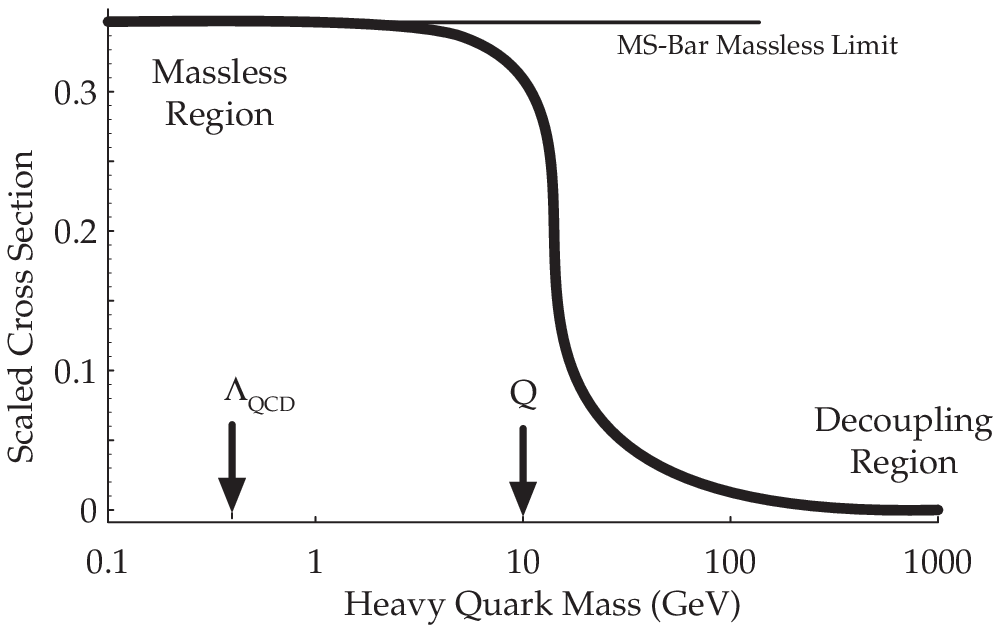}}
 \tightenlines
 \caption{\noindent 
 The scaled cross section for  DIS heavy quark production as a function of 
the quark mass $m_H$. 
 }
 \label{fig:xslimita}
\end{figure}
}
\def\figxslimitb
{
\begin{figure}[t]
 \epsfxsize=0.60\hsize  \centerline{\epsfbox{\dir 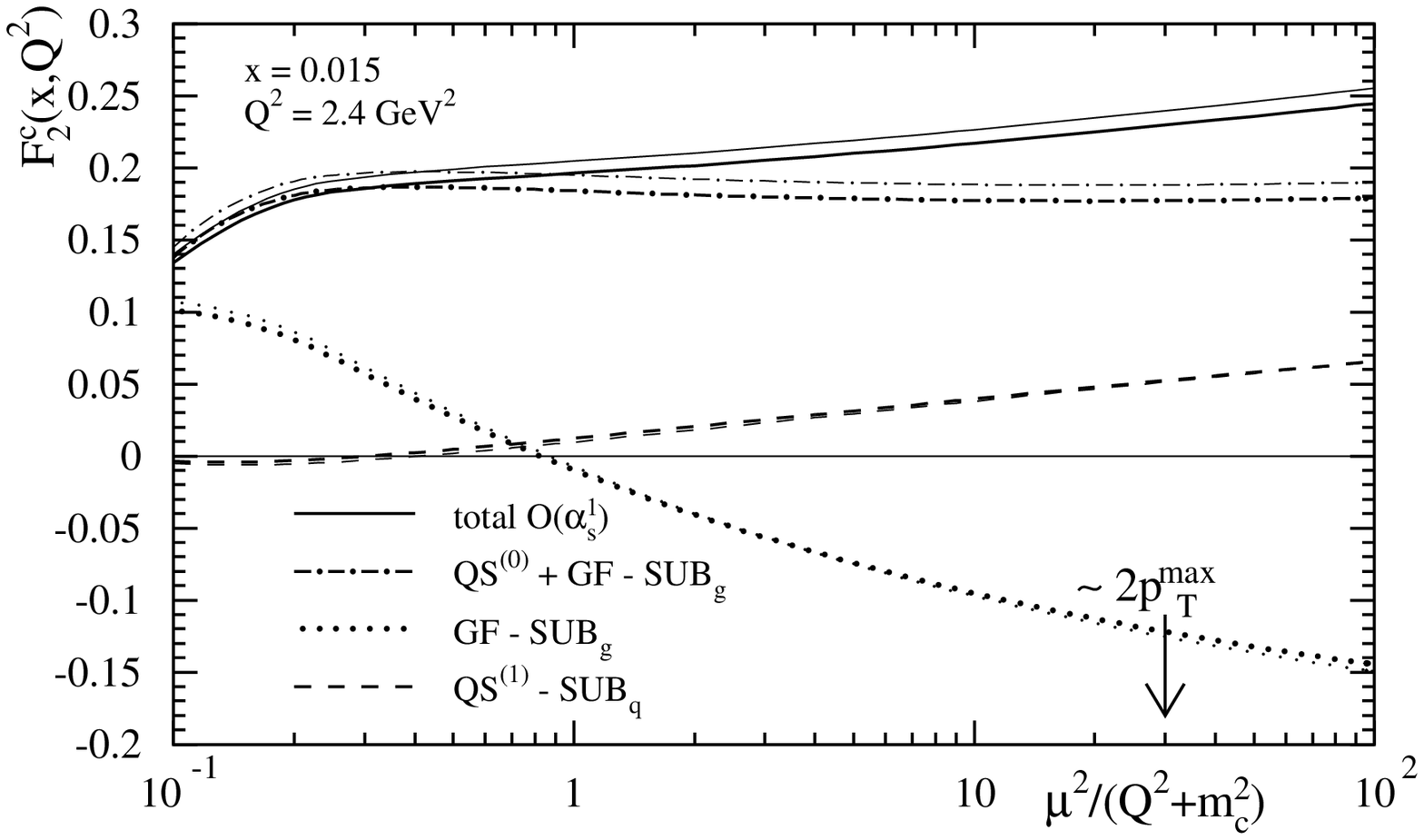}}
 \tightenlines
 \caption{\noindent 
  The contributions to DIS charged current inclusive $F_2^{charm}$ vs.\ $\mu$.
For each separate contribution, the thick lines are the \MSbar\ result ($m_s=0$), 
and the thin lines are the ACOT result with $m_s=0.5$ GeV.
 From Kretzer, Ref.~\protect\cite{kretzer}.
 }
 \label{fig:xslimitb}
\end{figure}
}
\def\figone 
{
\begin{figure}[t]
 \epsfxsize=0.90\hsize  \centerline{\epsfbox{\dir 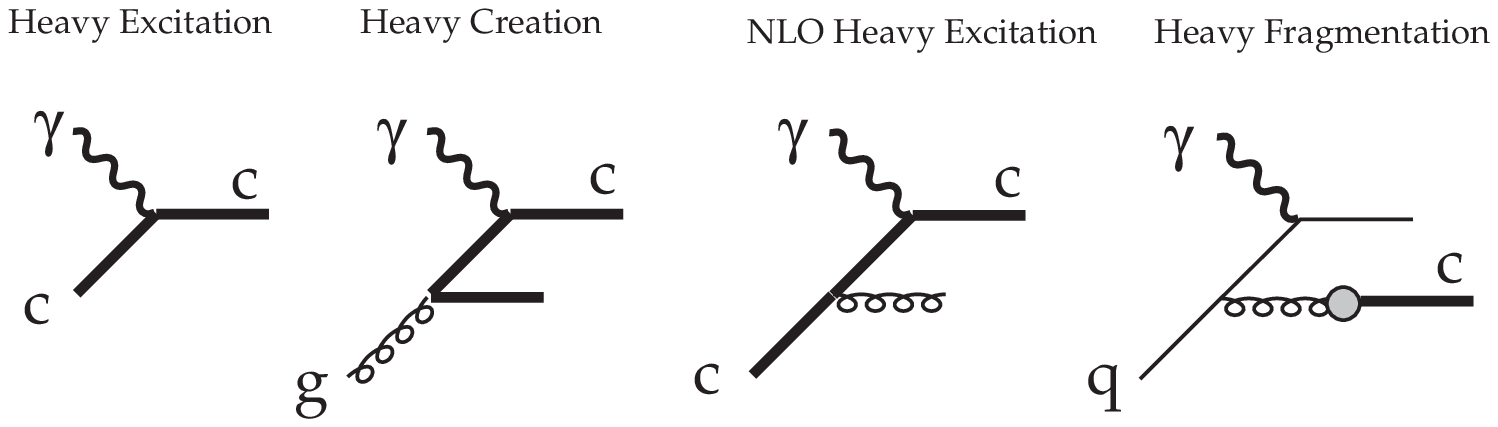}}
 \tightenlines
 \caption{\noindent 
Basic processes for DIS heavy quark production. 
 a) ${\cal O}(\alpha_s^0)$ flavor excitation: $\gamma + c \to c$;
 b) ${\cal O}(\alpha_s^1)$ flavor creation:   $\gamma + g \to c + \bar{c}$;
 c) ${\cal O}(\alpha_s^1)$ flavor excitation: $\gamma + c \to c + g$;
 d) ${\cal O}(\alpha_s^1)$ light-quark ($q$) fragmentation: $(\gamma +q \to q +g) \otimes (g \to c)$.
 }
 \label{fig:one}
\end{figure}
}
\def\figmassevl 
{
\begin{figure}[t]
 \epsfxsize=0.90\hsize \centerline{\epsfbox{\dir 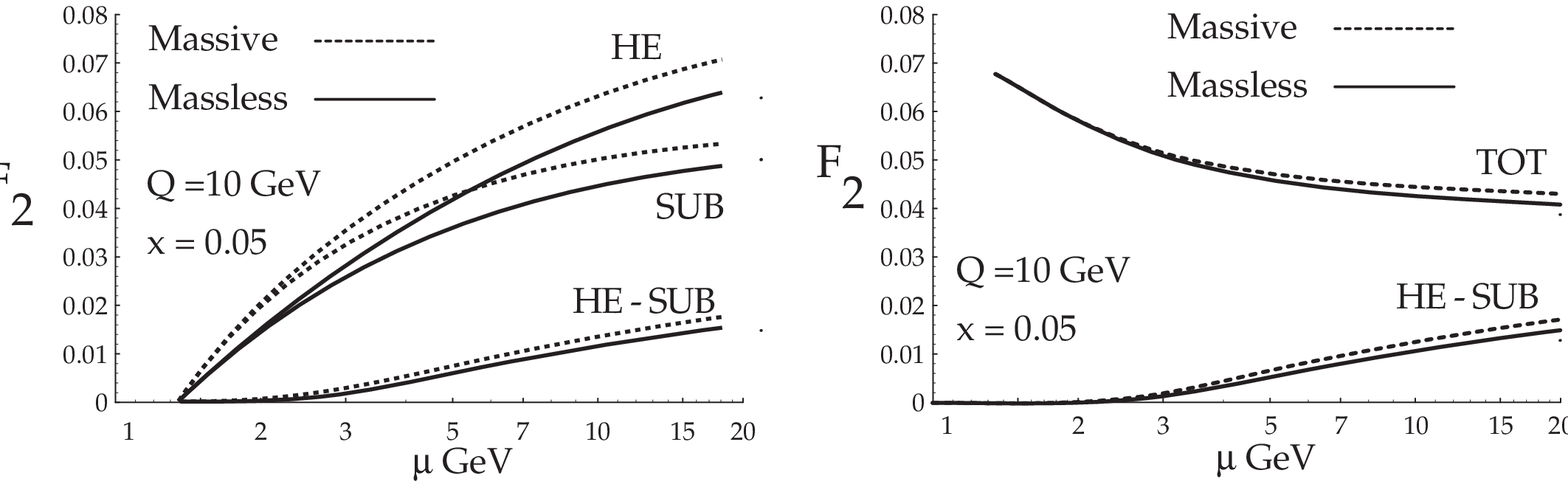}}
 \tightenlines
 \caption{\noindent 
 $F_2$ {\it vs.}\ $\mu$ for DIS c-production. 
 a) $F_2^{HE}$, $F_2^{SUB}$ and the difference $(F_2^{HE}-F_2^{SUB})$.
The  solid curves are for the mass-independent evolution scheme, 
and the   dashed  curves are for the mass-dependent evolution scheme.
 b) $F_2^{TOT}$ and $F_2^{HE}-F_2^{SUB}$. 
 The difference between the mass-independent evolution and 
mass-dependent evolution for $F_2^{TOT}$ is 
higher order and comparable or less than the $\mu$-variation.
  From   Ref.~\protect\cite{dis97oln}.
  }
 \label{fig:massevl}
\end{figure}
}
\def\figseligman
{
\begin{figure}[t]
 \epsfxsize=0.75\hsize  \centerline{\epsfbox{\dir 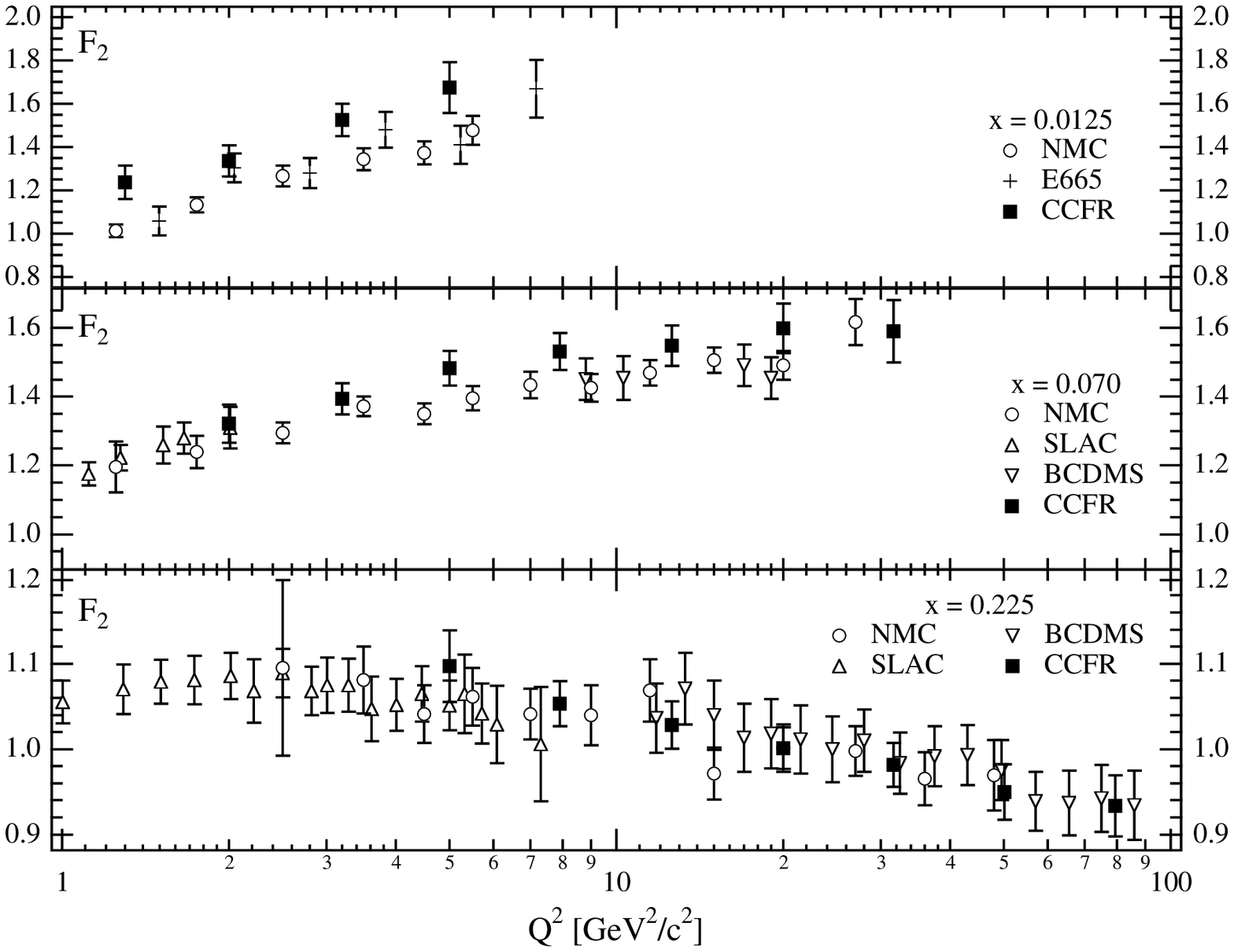}}
 \tightenlines
 \caption{\noindent 
 Comparison of $F_2$ from charged and neutral current DIS. 
 From Seligman, {\it et al.}, Ref.~\protect\cite{seligman}. 
 }
 \label{fig:seligman}
\end{figure}
}

\def\figtung
{
\begin{figure}[t]
 \epsfxsize=0.75\hsize  \centerline{\epsfbox{\dir figtung.eps}}
 \tightenlines
 \caption{\noindent 
 Comparison of H1 data in the small-$x$ region.
 From Lai and Tung, Ref.~\protect\cite{LaiTun97a}.
 }
 \label{fig:tung}
\end{figure}
}

\epsfverbosetrue

\def\gsim{\, \lower0.5ex\hbox{$\stackrel{>}{\sim}$}\, }
\def\lsim{\, \lower0.5ex\hbox{$\stackrel{<}{\sim}$}\, }
\def\MSbar{$\overline{{\rm MS}}$}
\def\eq#1{Eq.(\ref{eq:#1})}
\def\fig#1{Fig.~\ref{fig:#1}}

\begin{document}

\def\thefootnote{\alph{footnote}}
\def\thefootnote{\fnsymbol{footnote}}

\title{
\vspace*{-1in}
\hfill\parbox[t]{1.3in}
{\normalsize
                                                       hep-ph/9812270  \\
                                                       SMU-HEP/98-07    
}\\~\\~\\~\\ \bf   
Heavy Quark Production\footnote{%
 Presented at 4th Workshop on Heavy Quarks at Fixed Target (HQ 98), 
Fermilab, Batavia, IL, 10-12 Oct 1998.}
}

\author{Fredrick I. Olness}

\address{
Department of Physics, 
Southern Methodist University\\
Dallas, Texas 75275-0175\\
}

\maketitle

\begin{abstract}
 We  provide a brief overview of some current experimental and
theoretical issues of heavy quark production. 
\end{abstract}

\def\thefootnote{\fnsymbol{footnote}}
\def\thefootnote{\alph{footnote}}
\def\thefootnote{\arabic{footnote}}
\addtocounter{footnote}{-1}


\section*{Introduction}

The production of heavy quarks in high energy processes has become an
increasingly important subject of study both theoretically and
experimentally. 
The theory of heavy quark production in perturbative Quantum
Chromodynamics (pQCD) is more challenging than that of light parton (jet)
production because of the new physics issues brought about by the additional
heavy quark mass scale. The correct theory must properly take into account the
changing role of the heavy quark over the full kinematic range of the
relevant process from the threshold region (where the quark behaves like a
typical ``heavy particle'') to the asymptotic region (where the same quark
behaves effectively like a  massless parton).
 With steadily improving experimental data on a variety of processes
sensitive to the contribution of heavy quarks (including the direct
measurement of heavy flavor production), 
this is a very rich field for studying different aspects of the QCD theory 
including the problems of multiple scales, summation of large logarithms, 
subtleties of renormalization, and higher order corrections. 
 We shall briefly review a limited subset of these 
 issues.\footnote{%
For a recent comprehensive review, see: 
Frixione, Mangano, Nason, and Ridolfi, Ref.~\cite{FMNR97a}
}

\section*{The Factorization Theorem}

Perturbative calculations for heavy quark production 
are performed in the context of the factorization 
theorem expressed below in the commonly used form:
 \begin{eqnarray}
\sigma_{a \to c}
&=&
f_{a\to b}(x,\mu^2)  
\otimes 
\hat{\sigma}_{b \to c}(Q^2/\mu^2,M_H^2/\mu^2,\alpha_s(\mu)) 
+ {\cal O}(\Lambda^2_{QCD}/Q^2)
\label{eq:faci}
\end{eqnarray}
While the factorization was originally proven for massless quarks,\cite{CSS85} 
the theorem has recently been extended by Collins\cite{Collins97} 
to incorporate quarks of any mass, including ``heavy quarks."
(Note, we have explicitly retained the $M_H^2$ dependence in  $\hat{\sigma}$.)
 It is important to note that the corrections to the factorization 
are only of order $\Lambda^2_{QCD}/Q^2$, and not $M_H^2/Q^2$, 
{\it even for the case of general quark masses.}

The factorization theorem can also be expressed as a composition of 
$t$-channel two particle irreducible (2PI) 
amplitudes:\footnote{I must necessarily leave out many details here; 
for a precise treatment, see Collins\cite{Collins97}.}
 \begin{eqnarray}
\sigma_{a \to c}
&\simeq&
\hat{\sigma}_{b \to c}
\otimes 
f_{a\to b}
\simeq
\left[
C \cdot 
\frac{1}{1-(1-Z)K}
\right] 
\cdot  Z 
\ \cdot \ 
\left[ 
 \frac{1}{1-K} 
\cdot T 
\right]  
\label{eq:facii}
\end{eqnarray} 
 Here, 
C represents the graph for a hard scattering, 
K represents the graph for a rung, 
T represents the graph that couples to the target,
and
Z represents a collinear projection operator.
 The first term in brackets roughly corresponds to the hard scattering coefficient function $\hat{\sigma}$,
and the second term to the parton distribution function (PDF), $f$.
Note that these two terms only communicate through a collinear projection operator, Z. 
Part of the effort in generalizing the factorization theorem for the case of massive quarks involves
constructing the proper Z, and demonstrating that terms containing (1-Z) are power suppressed.
However, once Z is determined, \eq{facii}  yields an {\it all-orders} prescription for computing 
for both the hard scattering coefficient ($\hat{\sigma}$) and the parton distribution function ($f$). 
A calculation using this formalism was first performed by ACOT\cite{ACOT} for the case
of heavy quark production in deeply  inelastic scattering, and we now examine this process in detail.

\section*{Heavy Quark Production in DIS}

\figone

Several experimental groups\cite{HERA} have studied the semi-inclusive deeply  inelastic scattering (DIS)
process for heavy-quark production  $\ell_1 + N \to \ell_2 + Q + X.$ New data from HERA investigates the
DIS process in a very different kinematic range from that available at fixed-target experiments.  This
perception has changed the way that we compute the semi-inclusive  DIS heavy quark production. 
Traditionally, the heavy quark mass was treated as a large scale, and the number of active parton flavors
was fixed to be the number of quarks lighter than the heavy quark. In this scheme,  the perturbation
expansion begins with  the ${\cal O}(\alpha_s^1)$ heavy quark creation  fusion process 
$\gamma g \to c \bar{c}$, ({\it cf.}, \fig{one}b). We refer to this approach as the 
Fixed Flavor Number (FFN) scheme since the number of
flavors coming from parton distributions is fixed at three for 
charm production.\footnote{%
The necessary   diagrams have been computed to ${\cal O}(\alpha_S^2)$ 
 by Smith, van Neerven, 
and collaborators, {\it cf.}, Ref.~\cite{BSV}.
 }

\figxslimita

More recently, a Variable Flavor Number (VFN) scheme (ACOT \cite{ACOT}) 
has been proposed which includes the heavy quark
as an active parton flavor with
non-zero heavy quark mass. In this case,  the perturbation expansion begins with  the ${\cal
O}(\alpha_s^0)$ heavy quark excitation   process $\gamma c \to c$, ({\it cf.}, \fig{one}a).  The key
advantages of this scheme are:\cite{schmidt}
\begin{enumerate}
\item{}
 By incorporating the heavy quark into the parton framework, the composite scheme yields a result which is
valid from threshold to  asymptotic energies; in contrast, the FFN scheme contains unsubtracted  mass
singularities which will vitiate the perturbation expansion in the $m_c \to 0$ or $E \to \infty$ limit. 

\item{}
Because the composite scheme resums the large logarithms appearing in the FFN scheme into the parton
distribution functions, it includes  the numerically dominant terms of the ${\cal O}(\alpha_s^2)$
FFN scheme calculation in a ${\cal O}(\alpha_s^1)$ calculation.

\figxslimitb

\end{enumerate}
\noindent
In effect, the  VFN scheme subsumes the FFN scheme. 
 To illustrate this fact with a concrete calculation, in \fig{xslimita}, we plot the 
cross section for ``heavy" quark production as a function of the quark mass.\footnote{%
To be specific, we have computed single quark production for a photon exchange with $x=0.1$, 
$\mu=Q=10$ GeV, and the cross section is in arbitrary units.}
 This figure clearly shows the three important kinematic regions.
 1) In the massless region, where $m_H \ll Q$, the ACOT VFN result reduces precisely to the massless \MSbar\ result. 
 2) In the decoupling region, where $m_H \gg Q$, this ``heavy quark" decouples and its contribution vanishes. 
 3) In the transition region, where $m_H \sim Q$, this (not-so) ``heavy quark" plays an important dynamic role.
 While the FFN scheme is appropriate only when $m_H \sim Q$, we see that the VFN scheme is valid throughout the
full kinematic 
 range.\footnote{%
 Buza {\it et al.}, have determined the asymptotic form of the heavy quark coefficient 
functions which are then used to 
determine the threshold matching conditions between the three- and four-flavor
shemes,  Ref.~\cite{BSV}.
 Thorne and Roberts have a similar scheme with slightly different 
matching conditions, Ref.~\cite{thorne}.
}

This point is also illustrated in a calculation by Kretzer\cite{kretzer}
({\it cf.}, \fig{xslimitb}) which shows the partial contributions to 
the charged current $F_2^{charm}$.\footnote{%
 Kretzer and Schienbein have performed the first calculation of the 
${\cal O}(\alpha_S)$ quark initiated process
for general masses and general couplings,  Ref.~\cite{kretzer}.
 }
In this  figure, each line is actually a pair of lines: 
the thin lines represents the result for  $F_2^{charm}$  
using the  ACOT scheme with $m_s=0.5$ GeV,  
and the thick lines regularize the strange quark with the massless \MSbar\ prescription.
(The charm mass is, of course, retained.)
 The fact that these two calculations match so closely 
(particularly in comparison to the $\mu$-variation) indicates:
1) the ACOT scheme smoothly reduces to the desired massless \MSbar\ limit as $m_H\to 0$,
and 
2) for $m_H \lsim \Lambda_{QCD}$ we see that the quark mass no longer plays a dynamic role in the 
process and becomes purely a regulator.

\section*{Heavy Quarks and the Global PDF Analysis}

\figtung

 Recent precision data on $F_2$ and on $F_2^{charm}$ from HERA
indicate that the charm contribution can rise to 25\% of the total  $F_2$ at small-$x$.
 These results clearly imply the need to perform new global analyses to
account for the correct physics behind these measurements.
 Tung and Lai\cite{LaiTun97a} have repeated the CTEQ4M global analysis,\cite{CTEQ} but this time  
implementing the heavy quark leptoproduction within the ACOT formalism to obtain 
a CTEQ4HQ set of PDF's. 
 The deviation of CTEQ4HQ distributions from CTEQ4M are 
minimal, and  are most noticeable at small-$x$; 
 interestingly, the differences are larger for the light quarks than for the gluon and
charm.

The effect of these new PDF's and the comparison with data are shown in \fig{tung}. 
 The solid  curves show the CTEQ4M distributions convoluted with massless matrix elements. 
The dashed curves show the CTEQ4M distributions convoluted with massive matrix element;
while technically this is a mismatch of schemes, this comparison is useful to gauge 
the magnitude of the heavy quark effects, (which we observe are comparable to the 
experimental uncertainties). 
Finally, 
the dotted curves show the CTEQ4HQ distributions convoluted with massive matrix element.
 When a consistent scheme is used for both the matrix elements and the PDF's, the agreement
with data is excellent.  (This is as expected since this data was included in the fit.) 
 It is interesting to note that overall $\chi^2$ for CTEQ4HQ ($\chi^2$=1293) was slightly improved 
compared to the previous best fit CTEQ4M ($\chi^2$=1320) for 1297 data points. 
While this difference is small, we find it reassuring that the proper treatment of the
heavy quark mass resulted in an improved fit; 
particularly when compared with 
a 4-flavor FFN fit ($\chi^2$=1349) or 
a 3-flavor FFN fit ($\chi^2$=1380).

A recent re-analysis of the EMC data\cite{harris} concluded that there could be
an intrinsic charm component in the proton of $0.86\pm0.60\%$. 
 It would be  interesting to repeat this calculation in the context of a global analysis
using the VFN ACOT formalism to see if more recent data favor an intrinsic charm component.

\section*{Heavy Quarks and Extraction of $s(x)$}

\figseligman

A topic closely related to DIS charm production is the extraction of the 
strange quark distribution.\footnote{%
 For a comprehensive review, see  
 Conrad, Shaevitz, and  Bolton, Ref.~\cite{conrad}.
 }
 In principle, we can extract $s(x)$ by comparing DIS neutral and charged current
data. To leading order, we have:
 \begin{eqnarray}
\frac{F_2^{NC}}{F_2^{CC}}
&\simeq&
\frac{5}{18} \ 
\left\{ 
1 - \frac{3}{5} \frac{(s+\bar{s}) - (c+\bar{c}) + ...}{q+\bar{q}}
\right\}
\quad .
\label{eq:squark}
\end{eqnarray} 
 While the individual $F_2$ structure functions are measured precisely 
({\it cf.}, \fig{seligman}),\cite{seligman}  
this approach  is indirect in the sense that small uncertainties in the larger valence distributions
will magnify the uncertainty on the extracted $s(x)$.  

A direct method of obtaining $s(x)$ is to use the neutrino induced di-muon process:
$\nu_{\mu} N \to \mu^- c X$ with the subsequent decay $c\to s \mu^+ \nu_{\mu}$.
Here, the di-muon signal is directly related to the charm production rate, which 
goes via the process $W^+ s \to c$ at leading order. The method has the advantage
that the signal from the $s$-quark is not a small effect beneath the valence process.

A complete NLO experimental analysis was performed using the CCFR data set.\cite{CCFRcharm} 
 The recently collected data from the NuTeV experiment will provide an
opportunity to extend the precision of these investigations still
further.\cite{nutev}  Their high intensity sign-selected neutrino beam and the new
calibration beam allows for large improvement in the systematic uncertainty
while  minimizing statistical errors. (See the paper by T.~Adams, this meeting.\cite{adams})

\section*{Hadroprodution of Heavy Quarks}

\fighadroprod

We now turn to the hadroproduction of heavy quarks, and discuss 
how the method of ACOT\cite{ACOT,ost} is used to provide a dynamic role for the
heavy quark parton.
 We concentrate mostly on
b-production at the Tevatron for definiteness, and 
present typical results for $b$ quark production.\cite{FMNR97a,cdf,dzero}
(See the paper by A.~Zieminski, this meeting.\cite{Zieminski})
 \fig{hadroprod}a shows the scaled differential cross section  vs.\ $p_T$ for $b$
production at 1800 GeV for the leading order (LO) calculations. 
The heavy creation (HC) process\footnote{%
 In this section we let $g$ represent both gluons and light quarks, where applicable. 
 Therefore, the HC process described as $g g\to b \bar{b}$ also includes $q \bar{q}\to b \bar{b}$.}
 ($gg\to b \bar{b}$) represents the 
LO contribution to the fixed-flavor-number (FFN) scheme result. 
The heavy excitation (HE) process ($g b\to g b$) plus the HC term represents the 
LO contribution to the variable-flavor-number (VFN) scheme result. 
 The pair of lines for each result shows the effect of varying $\mu$.
In a similar manner, \fig{hadroprod}b  shows the total FFN and VFN results.\footnote{%
 The formidable calculations of the NLO $g g\to b \bar{b}$ process were computed by 
Nason, Dawson, and Ellis (Ref.~\cite{NDE}), and also by 
Beenakker {\it et al.},  (Ref.~\cite{Smithetal}). 
 These calculations were implemented in a Monte Carlo framework (including correlations) by 
 Mangano, Nason, and Ridolfi ,  (Ref.~\cite{MNR}).
  } 

Two interesting features are worth noting. 
1) Examining \fig{hadroprod}a we observe 
 the HE contribution is comparable to the HC one, in spite of the smaller $b$-quark PDF 
compared to the gluon distribution. Closer
examination reveals that two effects contribute to this unexpected result:
a larger color factor and the presence of
$t$-channel gluon exchange diagrams for the HE process, as compared to the HC process.
 2)  The LO-VFN (=HC+HE) contributions (\fig{hadroprod}a)
(tree processes) give a reasonable approximation to the 
full cross section TOT-VFN (\fig{hadroprod}b); thus, 
the NLO-VFN correction is relatively small. 
This is in sharp contrast to the familiar FFN scheme where the TOT-FFN term 
is more than twice as large as the LO-FFN (=HC).
 This is,  of course,
an encouraging result, suggesting that the VFN scheme heavy quark parton picture
represents an efficient way to organize the perturbative QCD series.

In \fig{hadroprod}a, we also observe that while the TOT-VFN result provides 
minimal $\mu$-variation for low $p_T$, the improvement is decreased for large $p_T$. 
This may be, in part, due to that fact that the TOT-VFN result shown here is missing
the NLO-HE process $g b\to g g b$ since this calculation, with masses retained, does not exist.
In a separate effort, Cacciari and Greco\cite{CacGre} have used a NLO fragmentation 
formalism to resum the heavy quark contributions in the limit of large $p_T$. 
This calculation effectively includes the massless limit of the $g b\to g g b$ contribution (omitted above);
the result is a decreased $\mu$-variation in the large $p_T$ region. 
Recently, this calculation has been merged with the massive FFN calculation
by Cacciari,  Greco, and Nason,  (Ref.~\cite{CGN}); 
the result is a calculation which matches the FFN calculation 
at low $p_T$, and  takes advantage of the NLO fragmentation formalism 
in the high $p_T$ region.

\section*{Massive vs. Massless Evolution}

\figmassevl

 In a consistently formulated pQCD framework incorporating non-zero
mass heavy quark partons, there is still the freedom to define
parton distributions obeying either  mass-independent or
mass-dependent evolution equations.
  With properly matched hard cross-sections, different choices merely
correspond to different factorization schemes, and they yield the
same physical cross-sections. 
 We demonstrate this principle in a concrete order $\alpha_s$
calculation of the DIS charm structure function.\cite{dis97oln}
In \fig{massevl}  we display
the separate contributions to $F_2^{charm}$  for both mass-independent 
and mass-dependent evolution. 
 The matching  properties   are best examined
by comparing the (scheme-dependent) 
heavy excitation  $F_2^{HE}$ and the subtraction $F_2^{SUB}$ contributions
of  \fig{massevl}a.

 We observe the following.
 1)~Within each scheme, $F_2^{HE}$ and $F_2^{SUB}$  are well
matched near threshold, {\it  cf.}, \fig{massevl}a.  
Above threshold, they begin to diverge, but the difference $(F_2^{HE}-F_2^{SUB})$, 
which contributes to $F_2^{TOT}$, is insensitive to the different schemes.  
 2)~It is precisely this matching of $F_2^{HE}$ and $F_2^{SUB}$ which ensures 
the scheme dependence of $F_2^{TOT}$ is properly of higher-order
in $\alpha_s$, ({\it  cf.}, \fig{massevl}b).  

This matching is not accidental, but simply a result of using a consistent renormalization 
scheme for both $F_2^{HE}$ and $F_2^{SUB}$.  
To understand this we expand these terms 
near threshold ($\mu \sim m_H$) where the $m_H/Q$ terms are relevant:
 \begin{eqnarray}
\sigma_{SUB} &=& 
{}^{R}f_{g/P} \otimes 
{}^{R}\widehat\sigma^{(1)}_{g\gamma^* \to c\bar{c}}  
= 
{}^{R}f_{g/P} \otimes
\frac{\alpha_s}{2 \pi}  
\int_{m_H^2}^{\mu^2}  \frac{d\mu^2}{\mu^2}  \ 
{}^{R}P^{(1)}_{g\to c} 
\otimes 
 \sigma^{(0)}_{c\gamma^* \to c} 
+ 0
\nonumber 
\\
\sigma_{HE} &\simeq& 
{}^{R}f_{c/P} \otimes {}^{R}\widehat\sigma^{(0)}_{c\gamma^* \to c} 
\simeq
{}^{R}f_{g/P} \otimes
\frac{\alpha_s}{2 \pi}  
\int_{m_H^2}^{\mu^2}  \frac{d\mu^2}{\mu^2}  \ 
{}^{R}P^{(1)}_{g\to c} 
\otimes 
 \sigma^{(0)}_{c\gamma^* \to c}
+ {\cal O}(\alpha_s^2)
\nonumber
 \end{eqnarray}
Here, the prescript $R$ specifies the renormalization scheme. 
 From these relations, it is evident that  $F_2^{HE}$ and $F_2^{SUB}$
will match to ${\cal O}(\alpha_s^2)$ so long as a consistent choice or renormalization scheme $R$ 
is made for the splitting kernels,
${}^{R}P^{(1)}_{g\to c}$.
 This is the key mechanism that compensates the different effects of
the mass-independent {\it vs.} mass-dependent evolution, and yields a
 $\sigma_{TOT}$ which is identical up to higher-order terms. 
 The lesson is clear:  the choice  of a mass-independent
\MSbar\ or a mass-dependent (non-\MSbar) evolution is  purely a choice
of scheme, and   becomes simply a matter of
convenience--{\it there is no physically new information gained
from the mass-dependent evolution.}

\section*{Conclusions}

 We have provided a brief overview of some current experimental and
theoretical issues of heavy quark production. 
 The wealth of recent heavy quark production data from both
fixed-target and collider experiments  will allow us to to extract  a
precision measurement of structure functions which can provide
important constraints  for searches of new physics at the highest
energy scales.
 As an important physical process involving the interplay of several
large scales, heavy quark production poses a significant challenge for
further development of QCD theory.

We thank 
J.C.~Collins,
R.J.~Scalise,
and
W.-K.~Tung
for valuable discussions,
and the Fermilab Theory Group for their kind hospitality during the 
period in which part of this research was carried out. 
 This work is supported
by the U.S. Department of Energy, and the
Lightner-Sams Foundation.



\begin{references}

\bibitem{FMNR97a}  %
S.~Frixione, M.~L.~Mangano, P.~Nason, and G.~Ridolfi,  hep-ph/9702287; 
M.~L.~Mangano, hep-ph/9711337.

\bibitem{CSS85}  J.~Collins, D.~Soper, and G.~Sterman, Nucl.\ Phys.\ {\bf B250}, 199 (1985).

\bibitem{Collins97}
J.~C.~Collins, Phys.\ Rev.\ {\bf D58}, 094002 (1998).

\bibitem{ACOT}  %
M.~A.~G.~Aivazis, J.~C.~Collins, F.~I.~Olness, and W.-K.~Tung, Phys.\ Rev.\
D {\bf 50}, 3102 (1994).


\bibitem{HERA}
H1 Collaboration (C.~Adloff {\it et al.}).
Z. Phys. C72, 593 (1996). 
\\
ZEUS Collaboration (J.~Breitweg {\it et al.}).  Talk given at
International Europhysics Conference on High-Energy Physics (HEP 97),
Jerusalem, Israel, 19-26 Aug 1997, N-645.

\bibitem{BSV}  %
E.~Laenen, S.~Riemersma, J.~Smith, W.L.~van Neerven. 
Phys.\ Rev.\ {\bf D49}, 5753 (1994);
M.~Buza, Y.~Matiounine, J.~Smith, and W.~L.~van~Neerven, hep-ph/9707263; 
hep-ph/9612398; %
M.~Buza and W.~L.~van~Neerven, Nucl.\ Phys.\ {\bf B500}, 301 (1997).

\bibitem{schmidt} 
 C.~Schmidt,   hep-ph/9706496;
J.~Amundson, C.~Schmidt, W.~K.~Tung, X.~Wang,  MSU
preprint, in preparation. 

\bibitem{kretzer}  %
 S.~Kretzer, e-Print   hep-ph/9808464, 
S.~Kretzer, I.~Schienbein, Phys.\ Rev.\ {\bf D58}, 094035 (1998).

\bibitem{thorne}
R.S. Thorne, R.G. Roberts,
Phys.\ Lett.\ {\bf B421}, 303 (1998);
R.S. Thorne, R.G. Roberts,
Phys.\ Rev.\ {\bf D57}, 6871 (1998).

\bibitem{LaiTun97a} 
H.~L.~Lai and W.-K.~Tung, Z.\ Phys.\ {\bf C74}, 463 (1997).
 The displayed figure is a reproduction of Fig.~2 from Lai and Tung,
 is copyrighted by Springer-Verlag, and used by permission. 


\bibitem{CTEQ}  %
H.~L.~Lai {\it et al.}, Phys.\ Rev.\ D {\bf 55}, 1280 (1997).

\bibitem{harris}  
B.W.~Harris, J.~Smith, R.~Vogt
Nucl.\ Phys.\ {\bf B461}, 181 (1996).

\bibitem{seligman} 
CCFR Collaboration (W.G.~Seligman et al.). hep-ex/9701017. 

\bibitem{conrad}
Janet~M.~Conrad, Michael~H.~Shaevitz, and Tim~Bolton. 
hep-ex/9707015 

\bibitem{CCFRcharm}  %
A.~O.~Bazarko {\it et al.}, Z.\ Phys.\ {\bf C65}, 189 (1995).

\bibitem{nutev}
NuTeV Collaboration: 
Jaehoon Yu     {\it et al}., hep-ex/9806030; 
K.S.~McFarland {\it et al}., hep-ex/9806013.

\bibitem{adams}
T.~Adams,{\it  Heavy Quark Production in Neutrino Deep-Inelastic Scattering.}
HQ'98, Fermilab, October 10-12, 1998.

\bibitem{ost} 
F.I.~Olness, R.J.~Scalise, Wu-Ki~Tung, hep-ph/9712494.
Phys.\ Rev.\ {\bf D59}, 014506 (1999).

\bibitem{cdf}  %
CDF Collaboration (F.~Abe {\it et al.}), Phys.\ Rev.\ D {\bf 50}, 4252 (1994); 
Phys.\ Rev.\ Lett.\ {\bf 75}, 1451 (1995).

\bibitem{dzero}
D0 Collaboration (S.~Abachi {\it et al.}),
Phys.\ Rev.\ Lett.\ {\bf 74}, 3548 (1995).

\bibitem{Zieminski}
A.~Zieminski,{\it B Production and Onium production at the Tevatron.}
HQ'98, Fermilab, October 10-12, 1998.

\bibitem{NDE}  %
P.~Nason, S.~Dawson, and R.~K.~Ellis, Nucl.\ Phys.\ {\bf B303}, 607 (1988);
{\bf B327}, 49 (1989); {\bf B335}, 260(E) (1990).

\bibitem{Smithetal}  %
W.~Beenakker, H.~Kuijf, W.~L.~van~Neerven, and J.~Smith, Phys.\ Rev.\ D {\bf 40}, 54 (1989); 
W.~Beenakker, W.~L.~van~Neerven, R.~Meng, G.~A.~Schuler, and J.~Smith,
Nucl.\ Phys.\ {\bf B351}, 507 (1991).

\bibitem{MNR}
M.~L.~Mangano, P.~Nason, and G.~Ridolfi, Nucl.\ Phys.\ {\bf B373}, 295
(1992).

\bibitem{CacGre}  
M.~Cacciari and M.~Greco, Nucl.\ Phys.\ {\bf B421}, 530 (1994).

\bibitem{CGN}
M.~Cacciari, M.~Greco, and P.~Nason, hep-ph/9803400, 
J.\ High Energy Phys.\ {\bf 05}, 007 (1998).

\bibitem{dis97oln}  F.~I.~Olness and R.~J.~Scalise, Phys.\ Rev.\ D {\bf 57},
241 (1998).

\end{references}
\end{document}